\documentclass[doublecol]{epl2}
\usepackage{amsmath,amssymb,eucal,graphicx,textcomp,mdwtab,dcolumn}

\newcommand{\df}{{d_{\text{f}}}}
\newcommand{\dw}{{d_{\text{w}}}}
\newcommand{\msd}{\delta r^2_\text{av}(t)}

\newcommand{\imsd}{\delta r^2_\infty(t)}

\renewcommand\vec\vect
\newcommand\diff\upd
\newcommand\expect[1]{\left\langle\vphantom{\big(}#1\right\rangle}

\newcommand{\eq}[1]{eq.~\eqref{eq:#1}}
\newcommand{\fig}[1]{fig.~\ref{fig:#1}}

\newcommand{\eg}{\textit{e.g., }}
\newcommand{\ie}{\textit{i.e., }}

\title{Cluster-resolved dynamic scaling theory and universal corrections for transport on percolating systems}
\shorttitle{Cluster-resolved scaling theory for transport on percolating systems}

\author{Axel Kammerer \and Felix H{\"o}f\/ling
\and Thomas Franosch}

\institute{
Arnold Sommerfeld Center for Theoretical Physics (ASC)  and Center for
NanoScience (CeNS),  \\ Fakult{\"a}t f{\"u}r Physik,
Ludwig-Maximilians-Universit{\"a}t M{\"u}nchen, Theresienstra{\ss}e 37,
80333 M{\"u}nchen, Germany
}

\abstract{
For percolating systems, we propose a universal exponent relation connecting the leading corrections to scaling of the cluster size distribution with the dynamic corrections to the asymptotic transport behaviour at criticality.
Our derivation is based on a cluster-resolved scaling theory unifying the scaling of both the cluster size distribution and the dynamics of a random walker. 
We corroborate our theoretical approach by extensive simulations for a site percolating square lattice and numerically determine both the static and dynamic correction exponents.}

\begin{document}

\maketitle

% \linenumbers

% \section{Introduction}

Anomalous, subdiffusive transport has been widely observed in biological systems, \eg in the cytoplasm~\cite{Golding:2006+Weiss:2004} and in cell membranes~\cite{Nicolau:2007}. %,Schwille:1999}.
These findings are attributed to an obstructed motion due to the crowded nature of cellular environments, and
they have been related to the subdiffusive motion established for percolating systems~\cite{Sung:2006+Saxton:1994}.
The experiments have yielded a wide spectrum of almost continuously changing exponents, but a mechanism generating such a continuum of fractal exponents has not been identified yet.
Although the observation windows covered between one and three decades in time, an alternative interpretation in terms of apparent power laws resulting from crossover phenomenona over large windows is conceivable.

For a continuum percolation model, it has been shown recently that the crossover from pure subdiffusion to normal diffusion extends over five decades in time~\cite{Lorentz_PRL:2006+Lorentz_JCP:2008}; in addition, the asymptotic behaviour is slowly approached and the large corrections cannot simply be ignored. Thus, it is of general interest to develop a systematic description of universal corrections to scaling in percolating systems.

In this Letter, we generalise the dynamic scaling theory for transport on percolation clusters~\cite{benAvraham:DiffusionInFractals,Stauffer:Percolation} to include the leading corrections. We derive a new exponent relation for dynamic corrections that are inherited from the static structure.
Let us start by introducing some notation and a summary of our main results.

% \subsection{Summary of the main results}

First, the cluster size distribution in the infinite lattice becomes fractal, $n_s \sim s^{-\tau}$, directly at the percolation threshold $p_c$ for large clusters, $s\to \infty$.
The Fisher exponent $\tau$ can be related to the fractal dimension $\df$ of the infinite cluster via $\tau=1+d/\df$,
provided the space dimension $d$ is smaller than the upper critical dimension $d_c=6$.
The deviations from the power law for $n_s$ are to leading order again determined by a power law
\begin{equation}
n_s(p_c) \simeq A s^{-\tau}  \left( 1 + Bs^{-\Omega}\right) \quad \text{for} \quad s\to \infty\,.
\label{eq:ns_correction}
\end{equation}
The correction exponent $\Omega>0$ is by renormalisation group arguments expected to be again universal, \ie it depends only on the dimension of the system, but not on the details of the lattice. Its role is to quantify how fast the critical manifold is approached and microscopic details become irrelevant.
We have determined the value of the correction exponent to $\Omega = 0.77 \pm 0.04$ by extensive computer simulations for a two-dimensional (2D) square lattice.

Second, transport of a random walker on the incipient infinite cluster at $p=p_c$ is anomalous, quantified by the power-law dependence of the mean-square displacement
$\delta r_\infty^2(t) \sim t^{2/\dw}$ for large times $t\to\infty$, where $\dw$ denotes the walk dimension. The approach to this power law is also non-analytic,
\begin{equation}
\delta r_\infty^2(t)  \simeq A_\infty t^{2/\dw} \left(1 +  C_\infty t^{-y} \right) \quad \text{for} \quad t\to \infty\, ,
\label{eq:rinftycrit}
\end{equation}
giving rise to the dynamic correction exponent $y>0$.

We shall introduce a cluster-resolved scaling theory for a generalised probability distribution incorporating the static and dynamic properties and derive the new exponent relation
\begin{equation}
y \dw = \Omega \df \, .
\label{eq:y_Omega}
\end{equation}
By computer simulations, we demonstrate that the assumptions underlying our scaling hypothesis are indeed fulfilled and that the measured correction exponents are compatible with the relation above. We have provided evidence for \eq{y_Omega} for $d=3$ before~\cite{Lorentz_PRL:2006+Lorentz_JCP:2008}, and we will test our theory here for $d=2$; universality allows to choose the simple square lattice for this purpose.
In Table~\ref{tab:exponents}, we have compiled some critical exponents for transport on percolation clusters in two and three dimensions along with the predictions from this work. 

\begin{table}
\newcolumntype{d}{D{.}{.}{1,7}}
\newcolumntype{s}{D{/}{/}{2,3}}
% version for bold columns, copied from dcolumn.pdf
\makeatletter\newcolumntype{E}[3]{>{\boldmath\DC@{#1}{#2}{#3}}c<{\DC@end}}\makeatother
\newcolumntype{b}{E{.}{.}{2,7}}
\begin{tabular}[C]{cdd|d}
\hlx{hhv[3]}
$d$ &
\multicolumn{1}{c}{2} &
\multicolumn{2}{c}{3} \\ \hlx{v[3]hv[3]}
$\df$ & \multicolumn{1}{s}{91/48^\text{a}} & \multicolumn{2}{d}{2.530(4)^\text{c}} \\
$\nu$ & \multicolumn{1}{s}{4/3^\text{a}} & \multicolumn{2}{d}{0.875(1)^\text{b}} \\
$\boldsymbol \Omega$ & \multicolumn{1}{b}{0.77(4)^\text{e}} & \multicolumn{2}{d}{0.64(2)^\text{b}} \\
\hlx{v[3]hv}
& &
\multicolumn{1}{c|}{lattice} & \multicolumn{1}{c}{continuum} \\ \hlx{vhv}
$\dw$ & 2.878(1)^\text{a} & 3.88(3)^\text{a} & 4.81(2)^\text{d} \\
$z$ & 3.036(1) & 5.07(6) & 6.30(3) \\
$\boldsymbol y$ &
\multicolumn{1}{b}{0.49(3)^\text{e}} &
\multicolumn{1}{b|}{0.42(2)^\text{e}} &
\multicolumn{1}{b}{0.34(2)^\text{e}} \\ \hlx{vhh}
\end{tabular}
\caption{Static and dynamic exponents for the leading and sub-leading critical behaviour. The uncertainty in the last digit is indicated in parentheses. Sources: (a) ref.~\cite{Grassberger:1998}, (b) ref.~\cite{Lorenz:1998}, (c) calculated from $\tau=2.186(2)$~\cite{Jan:1998}, (d) continuum percolation theory yields $\dw=\df+2/\nu$~\cite{Machta:1985+Halperin:1985,Lorentz_PRL:2006+Lorentz_JCP:2008}, and (e) this work.
The dynamic universality class does not split for $d=2$~\cite{Machta:1985+Halperin:1985,Lorentz_LTT:2007}.
The values for $z$ are derived from an exponent relation; the predictions for $y$ are based on our proposed relation, \eq{y_Omega}.}
\label{tab:exponents}
\end{table}

\section{Scaling of the cluster size distribution}
As the probability $p$ for a site to be occupied approaches a certain critical value $p_c$, clusters of all sizes emerge and the system becomes self-similar. In the infinite lattice, this is reflected by the power law tail of the distribution $n_s \sim s^{-\tau}$;
in a finite lattice of box length $L$, the largest cluster is expected to contain of the order of $L^{\df}$ sites.
Off the percolation threshold, the linear extension of the largest finite cluster $\xi$, also referred to as correlation length, diverges in the infinite lattice, $\xi \sim |\epsilon|^{-\nu}$, where $\epsilon :=(p-p_c)/p_c$ denotes the reduced distance to the critical point.
We now  introduce  a scaling assumption for $n_s$ that encompasses all the singularities as $s\to \infty, \epsilon \to 0$, and $L\to \infty$, including also an irrelevant scaling field $u$,
\begin{equation}\label{eq:ns_scaling}
n_s(\varepsilon, L)= L^{-d-\df} \mathsf{N}\big(s L^{-\df},\varepsilon L^{1/\nu}, uL^{-\omega}\big)\,.
\end{equation}
The first scaling variable $s L^{-\df}$ states that the size $s$ of a cluster should be compared with
the size $L^{\df}$ of the largest cluster in the box.
The second argument $\varepsilon L^{1/\nu}$ suggests that the system appears similar upon zooming in, $L \mapsto \lambda L$, provided the correlation length is rescaled accordingly, $\xi \mapsto \lambda \xi$.
The scaling field $u$ is assumed to be the leading irrelevant variable  in the renormalisation group sense~\cite{Cardy:ScalingRenormalization},
encoding how some microscopic details of the system fade out as the critical point is approached.
Since here the system is finite, $n_s$ is an analytic function of all its arguments, and since
the arguments $(s,\varepsilon, u)$ appear linearly inside the scaling function $\mathsf{N}$, the same property holds for $\mathsf{N}$.
The leading and next-to-leading behaviour is extracted by expanding $\mathsf{N}$
to first order in the irrelevant scaling field $u$,
\begin{equation}\label{eq:ns_expanded}
\begin{split}
n_s(\varepsilon, L) & = L^{-d-\df} \mathsf{N}_0\big(s L^{-\df},\varepsilon L^{1/\nu}\big) \\
 & \qquad + L^{-d-\df-\omega} \mathsf{N}_1\big(s L^{-\df},\varepsilon L^{1/\nu}\big)\,,
\end{split}
\end{equation}
with new analytic functions $\mathsf{N}_0,\mathsf{N}_1$,
and a factor $u$ has been absorbed in $\mathsf{N}_1$. A similar form of \eq{ns_expanded} was already given by Margolina \emph{et al.}~\cite{Margolina:1984}.

\section{Static corrections to scaling}

For large systems, $L \gg \xi$, the dependence on the box size may be eliminated, and  \eq{ns_expanded}  for the cluster size distribution can then be written as
\begin{equation}
n_s(\varepsilon)=s^{-\tau} \mathcal{N}^\pm_0\left( s \xi^{-\df}\right)
\left[ 1  + s^{-\Omega} \mathcal{N}^\pm_1\left(s \xi^{-\df}\right) \right]
\label{eq:ns_correction_xi}
\end{equation}
with new one-parameter scaling functions $\mathcal{N}^\pm_0(\cdot)$ for the leading behaviour; the superscript $\pm$ distinguishes the sign of $\varepsilon$. The correction is parametrised in terms of  $\mathcal{N}^\pm_1(\cdot)$ and the new correction-to-scaling exponent $\Omega:=\omega/\df$.
Specified to the critical point,  this yields
$n_s(\varepsilon=0) \simeq A s^{-\tau} \left(1 + Bs^{-\Omega}\right)$,
with the non-universal amplitudes $A := \mathcal{N}^\pm_0(0)$ and $B :=\mathcal{N}^\pm_1(0)$.

\begin{figure}
\includegraphics[width=\linewidth]{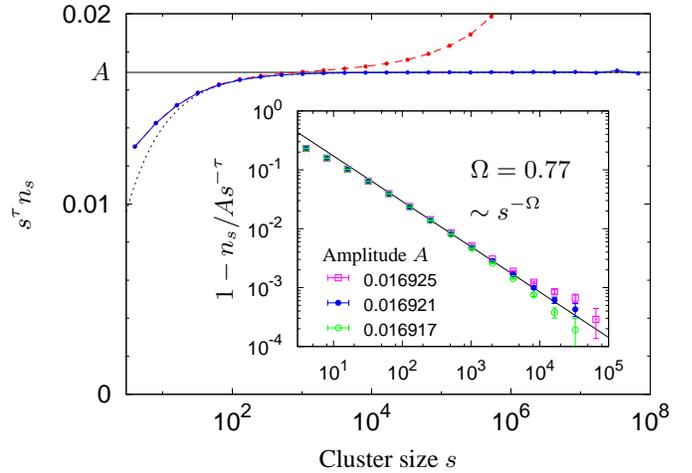}
\caption{Rectification of the cluster size distribution $n_s$ with the critical  power law $s^{-\tau}$ yields the amplitude $A$. Periodic (solid, blue) and free (broken, red) boundary conditions are compared. The dotted line indicates the leading correction to the asymptotic scaling, $n_s\simeq As^{-\tau}(1+Bs^{-\Omega})$. Inset: the deviation from the asymptotic law on a double-logarithmic plot follows a power law again with the universal correction exponent $\Omega=0.77$.}
\label{fig:ns_omega}
\end{figure}

A measurement of  the correction exponent $\Omega$ requires high quality data for the cluster size distribution $n_s$.
The deviation from leading order has to show an extremely small signal-to-noise ratio in order to follow the power law sufficiently long. This implies a precise knowledge about the leading order, $n_s\simeq A s^{-\tau}$, including the prefactor~$A$.
We have simulated 169,000 square lattices at the threshold,%
\footnote{The simulations required a total computation time of 90 hours on 100 cores of the AMD Opteron 285 2.6~GHz processor with 2~GB of RAM per core, which limited the size of the lattices.}
$p_c=0.592746$~\cite{Ziff:1992+Newman:2000}, using the multi-labelling algorithm by Hoshen and Kopelman~\cite{Hoshen:1976}.
We have used periodic boundary conditions and large lattices with a linear extend of $L=45{,}000$ sites to minimise the effects due to a finite simulation box~\cite{Rapaport:1985}. Figure~\ref{fig:ns_omega} shows a rectification of the obtained distribution $n_s$: the critical power law is followed over more than five decades in $s$. At small cluster sizes, the critical law is approached with another power law, $s^{-\Omega}$, as expected.
The figure also displays results for free boundary conditions, where pronounced finite-size effects, however, render a precise observation of the leading critical behaviour nearly impossible. 

In two dimensions, we take advantage of the Fisher exponent being known exactly, $\tau=187/91$. We have determined the amplitude of the leading order to $A=0.016921(4)$ by linear regression over various intervals in $10^4\leq s \leq 10^7$. Over these three decades, our data deviate from the leading power law by less than 1\textperthousand. The deviation from the asymptotic law is shown in the inset of \fig{ns_omega} for smaller cluster sizes, $s < 10^5$. On the double logarithmic plot, the data nicely follow a straight line again over more than two decades in the cluster size. But even with such accurate data, the precise extraction of the exponent is quite intricate. Since we need an asymptotic fit for large (but not too large) $s$, we varied the fit interval for the linear regression in $10^2\leq s \leq 10^4$; this procedure, however, seems to underestimate the exponent slightly in our case. A more reliable approach for an asymptotic fit is to do another rectification, now of the correction: the best convergence of $s^\Omega(n_s / A s^{-\tau}-1)\to B$ was found for $\Omega=0.77$ with the above value of $A$, yielding $B=-1.01$. The result slightly depends on $A$; varying the amplitude within the above uncertainty, we have obtained error bounds on the correction exponent,  $\Omega=0.77\pm0.04$.

Universality of $\Omega$ suggests it is independent of the lattice. Indeed, our result is fairly compatible with the result for the triangular lattice~\cite{Rapaport:1986}, where $\Omega$ was found to be between $0.71$ and $0.74$; we attribute the discrepancy to the limited observation window for the correction, $s<10^3$, 22~years ago.

\section{Dynamics of a random walker}

Transport on percolating systems is expected to be anomalous since a random walker (``the ant'') has to explore the self-similar  structures of the ramified clusters.
The frozen disorder leads to subdiffusive motion on the infinite cluster directly at the percolation threshold. An independent critical exponent $\dw>2$, referred to as walk dimension, characterises
the restricted mean-square displacement (MSD) defined as
$\delta r^2_\infty(t) := \expect{ [ \vec R(t)-\vec R(0) ]^2 \,|\,\vec R(0)\in\mathcal{C}_\infty} \sim t^{2/\dw}$,
where the average is taken over different realisations of the disorder and $\vec R(t)$ denotes the trajectory of the ant;
some values of $\dw$ are listed in Table~\ref{tab:exponents}.
If walkers on all clusters are considered, the exponent is modified since the average of clusters of all sizes is weighted with the fractal cluster size distribution~\cite{benAvraham:1982+Gefen:1983};
the unrestricted MSD follows $\delta r^2_\text{av}(t)\sim t^{2/z}$ with the dynamic exponent
$z:=2 \dw/(2+\df-d) > \dw$.
Away from the critical point, the long-time behaviour is either diffusive, $\delta r^2_\text{av}(t) \simeq 2d D t$, for $p> p_c$ or localised, $\delta r^2_\text{av}(t\to \infty) = \ell^2$,  below the transition, $p< p_c$.
The diffusion coefficient $D$ and the localisation length $\ell$ exhibit power law behaviour in~$\epsilon$.

\subsection{Cluster-resolved scaling}
The statistical information of the transport dynamics is encoded in the van Hove self-correlation function $G(\vec r,t):=\expect{\delta(\vec R(t)-\vec R(0)-\vec r)}$, \ie the probability that the ant has been displaced by a vector $\vec r$ in time~$t$.
Here  we suggest  a  cluster-resolved scaling theory for the van Hove function including the leading correction to scaling; in particular, we will derive a relation between static and dynamic exponents. The principle idea is that at the critical point all clusters resemble each other up to proper rescaling and therefore induce similar dynamics after readjusting the
time scale. To take advantage of the self-similarity, we generalise the van Hove function to also resolve the cluster size,
\begin{equation}
G_s(\vec r,t):=\expect{\delta(\vec R(t)-\vec R(0)-\vec r)|\vec R(0)\epsilon\, \mathcal{C}_s}\,,
\end{equation}
where $\mathcal{C}_s$ denotes a cluster of size $s\leq\infty$.
After averaging, the information on the detailed structure of the cluster is lost, and we keep only its size as indicator for the dynamics. The ant is trapped for finite $s$, whereas for $s=\infty$ the infinite cluster allows for long-range transport.
Let us introduce also the joint probability $P_s$ that the ant is on a cluster of size $s$ and has moved a vector $\vec r$ in a given time $t$,
as well as a corresponding probability $P_\infty$ for the infinite cluster~\cite{Hoefling:PhD_thesis},
\begin{subequations}
\begin{align}
P_s(\vec r,t;\varepsilon)&:=sn_s(\varepsilon)\,G_s(\vec r,t;\varepsilon)\,,
\label{eq:Ps} \\
P_\infty(\vec r,t;\varepsilon)&:=P(\varepsilon)\,G_\infty(\vec r,t;\varepsilon)\,;
\label{eq:Pinf}
\end{align}
\end{subequations}
$P(\varepsilon)$ is the probability that a given occupied site belongs to the infinite cluster.
We now postulate a scaling relation for the joint probabilities in the same spirit as for \eq{ns_scaling},
\begin{subequations}{\belowdisplayskip=0pt
\begin{multline}
P_s(\vec r,t;\varepsilon,L)  = \\
\hfil L^{-2d} \,\mathsf{P}_\text{F}\big(r/L, t L^{-\dw}, s L^{-\df}, \epsilon L^{1/\nu}, u L^{-\omega}\big) \, ,
\label{eq:Ps_scaling} 
\end{multline}}{\abovedisplayskip=0pt
\begin{multline}
P_\infty(\vec r,t;\varepsilon,L)  =  \\
\hfil L^{\df-2d} \,\mathsf{P}_\infty\big(r/L, t L^{-\dw}, \epsilon L^{1/\nu}, u L^{-\omega}\big) \, ,
\label{eq:Pinf_scaling}
\end{multline}}%
\label{eq:Ps+inf_scaling}%
\end{subequations}
where the finite size $L$ of the system has been explicitly considered. (For a finite system, the infinite cluster is defined as the largest cluster of a given percolating lattice.)
We have assumed that the leading irrelevant scaling variable $u$ is the same as in the static case.
Since the system is finite, $\mathsf{P}_\text{F}$ and $\mathsf{P}_\infty$ are analytic functions in all their arguments.
The choice of the arguments is motivated by the well-known scaling relations for the linear extent of walks (exponent $ \dw$), the cluster size ($\df$), and the correlation length ($\nu$). The prefactor ensures consistency with the scaling form for the cluster size distribution, \eq{ns_scaling}.
Since the van Hove functions $G_s$ are normalised, $s n_s$  is obtained by integrating \eq{Ps} over all $\vec r$, compatible with the scaling hypothesis for $n_s$. Integration of \eq{Pinf_scaling} yields the familiar result for the strength of the infinite cluster, $P\sim|\epsilon|^\beta$, with exponent $\beta=\nu(d-\df)$ as $L\to\infty$.
To simplify notation, we suppress the dependence of the scaling functions on $u$ in the remainder of this section.

The limit of infinite system size eliminates the dependence on $L$,
\begin{subequations}
\begin{align}
P_s(\vec r,t;\varepsilon)  &=  R_s^{-2d}\mathcal{P}_\text{F}\big(r/R_s, t R_s^{-\dw}, \varepsilon R_s^{1/\nu}\big) \, , \\
P_\infty(\vec r,t;\varepsilon)  &=  \xi^{\df-2d}\mathcal{P}_\infty\left(r/\xi, t \xi^{-\dw}\right) \, ,
\end{align}
\end{subequations}
where in the new scaling function $\mathcal{P}_\text{F}$, we have replaced the cluster size $s$ by the linear extent of $s$-clusters, $R_s \sim s^{1/\df}$, which serves as reference length scale.
Note that $\mathcal{P}_\text{F}$ is still analytic in all arguments. For the van Hove functions, this implies
a scaling form
\begin{subequations}
\begin{align}
G_s(\vec r,t;\varepsilon)  &=  R_s^{-d}\mathcal{G}_\text{F}\big(r/R_s, t R_s^{-\dw}, \varepsilon R_s^{1/\nu}\big) \, , \\
G_\infty(\vec r,t;\varepsilon)  &=  \xi^{-d}\mathcal{G}_\infty\left(r/\xi, t \xi^{-\dw}\right) \, .
\end{align}
\end{subequations}
The second moments define the cluster-resolved MSDs and possess the scaling forms
\begin{subequations}
\begin{align} \label{eq:msds_scaling}
\delta r_s^2(t;\varepsilon) &=
t^{2/\dw}\delta\mathcal{R}^2_\text{F}\big(t R_s^{-\dw}, \varepsilon R_s^{1/\nu}\big)\,, \\
\label{eq:msdinf_scaling}
\delta r_\infty^2(t;\varepsilon) &=
t^{2/\dw}\delta\mathcal{R}^2_\infty\left(t \xi^{-\dw}\right)\,.
\end{align}
\label{eq:cr_msd_scaling}%
\end{subequations}
Note, that most clusters are much smaller than the correlation length, $R_s\ll \xi$,  implying
$|\varepsilon|  R_s^{1/\nu} \ll  1$. For these clusters, $\delta r_s^2(t;\varepsilon)$ is insensitive to the distance to the critical point $\varepsilon$, and thus \eq{msds_scaling} essentially constitutes a one-parameter scaling prediction, which we will discuss below.
The second relation, \eq{msdinf_scaling}, is the familiar result for a random walker restricted to the infinite cluster.
% Similar reasoning for the infinite cluster leads to the scaling prediction the MSD
% \begin{equation}\label{eq:msdinf_scaling}
% \delta r_\infty^2(t;\varepsilon) =t^{2/\dw}\delta\tilde{r}^2(t \xi^{-\dw} ).
% \end{equation}

\begin{figure}
  \includegraphics[width=\linewidth]{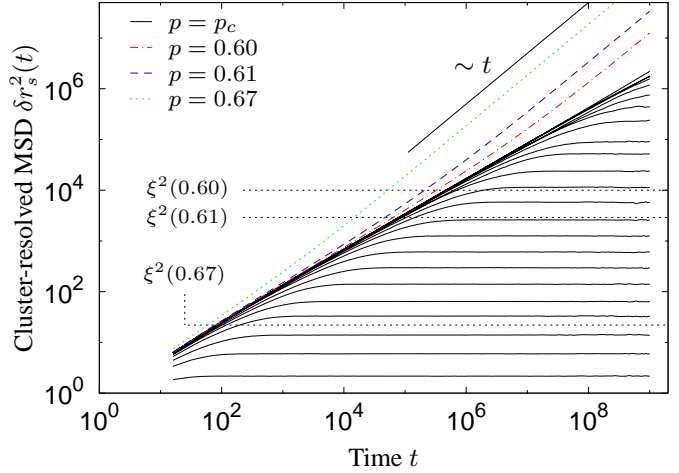}
  \caption{Cluster-resolved MSDs $\delta r^2_s(t)$ at the threshold (solid lines); $\imsd$ is shown for several densities above the threshold. Distances are measured in units of the lattice spacing and time refers to the number of hopping attempts.}
  \label{fig:clustermsd}
\end{figure}

\begin{figure}
  \includegraphics[width=\linewidth]{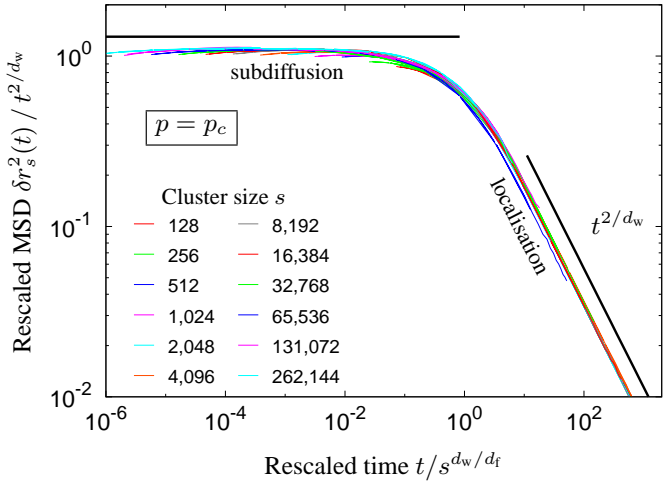}
  \caption{Cluster-resolved data from \fig{clustermsd} rescaled with $s^{1/\df}$, which is proportional to the radius of gyration $R_s$.}
  \label{fig:msd_scaling}
\end{figure}

We have generated a total of more than 1000~trajectories
on 20~square lattices with a linear extent of $L=8{,}000$ sites for several densities $p$ each. Periodic boundaries were used to minimise finite-size effects, the starting points of the ant were chosen at random, and the size of the clusters was determined using a union-find algorithm~\cite{Newman:2001}.%
% \footnote{For 1,500 trajectories 200 hours were required using 10 processor cores of type Quad Core Intel\textsuperscript{\textregistered}  Xeon\textsuperscript{\textregistered}  CPU X5365 3.0\,GHz.}

The cluster-resolved MSDs $\delta r^2_s(t)$ at the threshold  are exhibited in \fig{clustermsd};
all of them follow the same power law $t^{2/\dw}$ before they saturate.
The scaling prediction \eq{msds_scaling} implies that the respective long-time limits yield the radii of gyration and grow as $R_s \sim s^{1 / \df}$. The crossover time from subdiffusion to localisation is expected to scale as $t_\times \sim R_s^{\dw}$.
Figure~\ref{fig:msd_scaling} demonstrates that our data exhibit this scaling remarkably well; they bridge the crossover region from anomalous transport to localisation, which extends over three orders of magnitude in time.
We have also included the MSD on the infinite cluster $\imsd$ above the threshold in \fig{clustermsd}.
The crossover to diffusive transport occurs once the MSD is of the order of $\xi^2$, \ie right at the length scale above which the system looks homogeneous, or in terms of times, $t_\times \sim \xi^{\dw}$,
in consistency with \eq{msdinf_scaling}.
The scaling prediction is illustrated in \fig{msd_epsilonscaling}, corroborating an earlier study~(fig.~6.2 in ref.~\cite{benAvraham:DiffusionInFractals}).
In summary, the observed scaling behaviour validates the notion of self-similarity also for the structures of the clusters and the transport on them, provided the system is close to the critical point.

\subsection{All-cluster-averaged MSD}
The all-cluster-averaged MSD $\delta r^2_\text{av}(t)$ is calculated from the data for the cluster-resolved MSDs and the weights $s n_s$ and $P$,
\begin{equation}\label{eq:nsrs}
\delta r^2_\text{av}(t;\varepsilon)=\sum_s sn_s(\varepsilon)\,\delta r^2_s(t) + P(\varepsilon)\,\delta r^2_\infty(t)\,.
\end{equation}
Using  the scaling predictions eqs.~\eqref{eq:ns_correction_xi} and \eqref{eq:cr_msd_scaling}, one easily derives
$\delta r^2_\text{av}(t;\varepsilon) =t^{2/z}\widetilde{\delta \mathcal{R}}{}_\pm^2(t\xi^{-\dw})$,
where the dynamic exponent $z$ emerges.
At criticality, the infinite cluster has zero weight and the finite clusters contribute an infinite hierarchy of cluster sizes,
resulting in an unbounded MSD, diverging as $\delta r^2_\text{av}(t)\sim t^{2/z}$.
Below the threshold, $\delta r_s^2(t\to\infty ) \sim R_s^2$ implies that the long-time limit of the averaged MSD measures the mean-square cluster size, $\delta r^2_\text{av}(t\to\infty) =: \ell^2$, where $\ell \sim |\varepsilon|^{-\nu + \beta/2}$. Consistency with the dynamic scaling for $\delta r^2_\text{av}(t)$ requires $\ell^z \sim \xi^{\dw}$, allowing to write the scaling form more transparently,
\begin{equation}\label{eq:msd_scaling}
\delta r^2_\text{av}(t) = t^{2/z}\delta \mathcal{R}{}^2_\pm\left(t \ell^{-z}\right)\,.
\end{equation}
On large time scales when $\delta r^2_\text{av}(t)\gg \xi^2$, merely the infinite cluster contributes to the MSD and transport becomes diffusive. With the ansatz $\delta r^2_\text{av}(t)\sim D t$, \eq{msd_scaling} dictates that the diffusion coefficient $D$ vanishes at the threshold as $D\sim |\varepsilon|^\mu$ where $\mu=\nu(\dw-2)+\beta$.

\begin{figure}
\includegraphics[width=\linewidth]{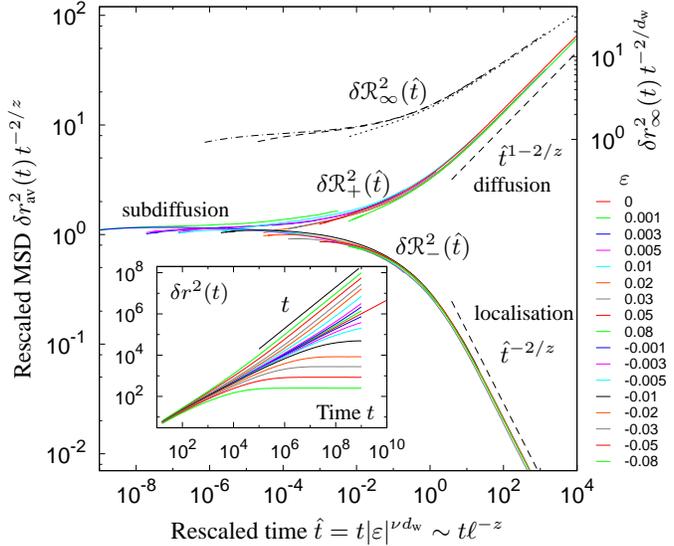}
\caption{All-cluster-averaged MSDs $\delta r^2_\text{av}(t;\epsilon)$ close to the threshold. While the inset shows the raw data, rescaling according to \eq{msd_scaling} collapses all data on the scaling functions $\delta\mathcal{R}^2_\pm(\dot)$. Right axis: rescaled MSDs on the infinite cluster, $\delta r^2_\infty(t;\epsilon)$, from \fig{clustermsd}, following \eq{msdinf_scaling}.}
\label{fig:msd_epsilonscaling}
\end{figure}

Close to the percolation threshold, our numerical results for the MSD $\delta r^2_\text{av}(t;\epsilon)$ follow the critical law, $\delta r^2_\text{av}(t)\sim t^{2/z}$, until the curves fan out to either diffusive or localised behaviour (inset of \fig{msd_epsilonscaling}).
We have rescaled the simulated data according to \eq{msd_scaling} and find excellent data collapse, see \fig{msd_scaling}. The crossover region covers approximately four nontrivial decades in rescaled time.
Let us emphasise that to achieve a data collapse of similar quality for the 3D Lorentz model, it was necessary to consider explicitly the leading corrections to scaling~\cite{Lorentz_PRL:2006+Lorentz_JCP:2008}.

\section{Corrections to dynamic scaling}
The leading dynamic corrections to scaling are derived from the scaling form of the joint probabilities for finite and infinite clusters, eqs.~\eqref{eq:Ps+inf_scaling}, along the same lines as for the cluster size distribution.
For the MSD on the infinite cluster this implies $\delta r_\infty^2(t;\varepsilon) =t^{2/\dw}\delta\mathsf{R}^2_\infty(t \xi^{-\dw}, u \xi^{-\omega} )$. Expanding to first order in the irrelevant scaling variable $u$ yields
\begin{equation}
\delta r_\infty^2(t)=t^{2/\dw}\delta \mathcal{R}_\infty^2\left(t\xi^{-\dw}\right) \left[1+t^{-y}\Delta_\infty\left(t\xi^{-\dw}\right)\right],
\label{eq:msdinf_correction}
\end{equation}
where the correction is quantified by the new exponent $y=\omega/\dw$ and a new scaling function $\Delta_\infty(\cdot)$. Specifically at the critical point, the cluster-resolved scaling hypothesis predicts $\delta r_\infty^2(t)= A_\infty t^{2/\dw}\left(1+C_\infty t^{-y}\right)$ with the non-universal amplitudes $A_\infty:=\delta \mathcal{R}^2_\infty(0)$ and $C_\infty:=\Delta_\infty(0)$, compare \eq{rinftycrit}. Thus, we have derived the exponent relation, $y\dw =\omega=\Omega\df$, between the dynamic and static correction exponents $y$ and $\Omega$, mentioned in the introduction.

A similar chain of arguments for the MSD on the finite clusters extends \eq{msds_scaling} at criticality to
\begin{equation}
\delta r^2_s(t)=t^{2/\dw}\delta \mathcal{R}^2_\text{F}\left(tR_s^{-\dw}\right)
\left[1+t^{-y}\Delta_\text{F}\left(t R_s^{-\dw}\right)\right].
\label{eq:msds_correction}
\end{equation}
Observing that the MSDs on a finite and the infinite cluster are indistinguishable at not too long time scales, $t\ll R_s^\dw,\xi^\dw$, we infer $\delta\mathcal{R}^2_\text{F}(0)=A_\infty$ and $\Delta_\text{F}(0)=C_\infty$.
Calculating the all-cluster average, \eq{nsrs}, from eqs.~\eqref{eq:ns_correction_xi},\eqref{eq:msdinf_correction}, \eqref{eq:msds_correction} leads to
\begin{equation}
\delta r^2_\text{av}(t)=t^{2/z}\delta \mathcal{R}_\pm^2\left(t\ell^{-z}\right)
\left[1+t^{-y}\Delta_\pm\left(t\ell^{-z}\right)\right],
\end{equation}
where the scaling variable can equivalently be written as $t\xi^{-\dw} \propto t\ell^{-z}$.
We have suggested such a form already for the 3D Lorentz model~\cite{Lorentz_PRL:2006+Lorentz_JCP:2008}, and here we provide the basis for this ansatz. In particular, we shall test the prediction at the critical point,
\begin{equation}
\delta r^2_\text{av}(t) \simeq A_\text{av} t^{2/z}\left(1+C_\text{av}t^{-y}\right)
\quad \text{for} \quad t\to\infty\,.
\end{equation}

\begin{figure}
\includegraphics[ width=\linewidth]{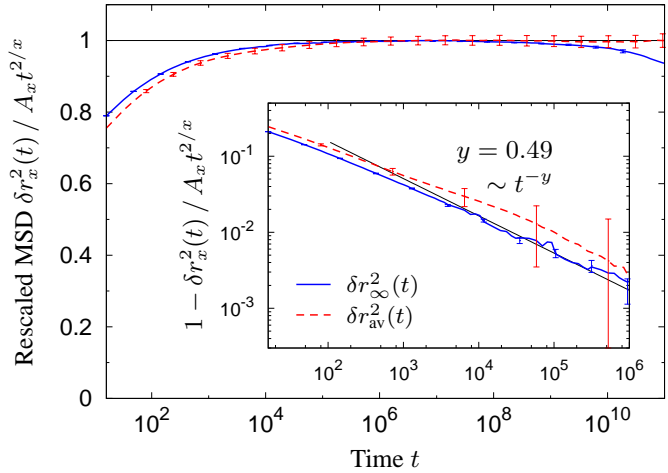}
\caption{Rectified MSDs restricted to the infinite cluster (solid, blue) and averaged over all clusters (broken, red) at the critical density. Data for $\imsd$ are rescaled with $A_\infty t^{2/\dw}$ and for $\msd$ with $A_\text{av} t^{2/z}$.
Inset: a double-logarithmic plot of the relative deviations from the asymptotic behaviour shows that the leading corrections indeed follow the predicted power law.}
\label{fig:ycombined}
\end{figure}

For a numerical analysis of the dynamic corrections to scaling, we have simulated 4,070 trajectories of $10^{11}$~steps on periodic square lattices with a linear extent of $L=90{,}000$ sites at the critical density;%
\footnote{Each trajectory required a computing time of about 19 hours on a single core of the AMD Opteron 8218 2.6~GHz processor.}
the starting points were chosen at random on arbitrary clusters. To increase computational efficiency, we utilised the statistical independence on the bit-level of the random number generator GFSR4~\cite{Ziff:1998}: calculating one random 32-bit integer allowed the realisation of 16 hopping attempts.
Figure~\ref{fig:ycombined} shows the rectified MSD $\delta r^2_\text{av}(t) / A_\text{av} t^{2/z}$ at the threshold; a fit for the amplitude yields $A_\text{av}=1.204(1)$. The leading asymptotic behaviour is very well captured over 6 decades in time with the exponent $z$ obtained from conductivity measurements, see table~\ref{tab:exponents}. On a double-logarithmic scale, the deviations from the leading behaviour are compatible with a power law decay with exponent $y=0.49$. There is still a relatively large, overall statistical error in our data of about 1\% since the heterogeneous average over clusters of all sizes converges quite slowly.
Better convergence is expected if the random walkers are restricted to the infinite cluster alone. We have generated 2,266 trajectories
% \footnote{The computing time was about 23 hours per trajectory.}
at criticality on the largest cluster of periodic lattices with $L=13{,}500$, which allowed a measurement of $\delta r^2_\infty(t)$ with a precision of 1\textperthousand{} for $t\lesssim 10^8$ steps.  Rescaled with the asymptotic law $A_\infty t^{2/\dw}$ where $A_\infty=1.152(3)$, the data converge to a plateau within 3\textperthousand{} for a decade in time; for times longer than $t\gtrsim 10^7$, the data start to deviate from the plateau, we have checked that this is a finite-size effect. Nevertheless, the approach to the plateau at shorter time scales is well captured with our predicted power law, $t^{-y}$ with $y=0.49$, over two decades in time, see inset of \fig{ycombined}. It not only confirms the power law form of the leading corrections to scaling, but also the derived exponent relation between $\Omega$ and $y$, \eq{y_Omega}.

\section{Conclusions}
We have derived an exponent relation connecting the leading corrections to scaling of the static cluster structure and the transport dynamics. The derivation relies on a cluster-resolved scaling theory unifying structure and dynamics,
which has been corroborated in detail by simulations for the square lattice.

The corrections to scaling discussed here have their origin in the same leading irrelevant scaling field $u$ for the statics and dynamics.
Yet, the motion of a random walker on the fractal clusters introduces new, genuinely dynamic scaling fields, one relevant and infinitely many irrelevant ones. In our approach, we have anticipated that the \emph{leading} irrelevant scaling field is still provided by the static field $u$, in agreement with our simulations.
Analytic deformations of the scaling fields typically lead to analytic corrections with integral relative exponents~\cite{Cardy:ScalingRenormalization}. Here, we find $\Omega,y<1$ and conclude that the universal corrections dominate over the analytic contributions.

Finally, our analysis suggests to interpret experiments on anomalous diffusion in disordered environments in terms of universal exponents, including leading corrections to the asymptotic behaviour, rather than fitting with phenomenological exponents. Percolation provides one possible universality class for such an approach.

\acknowledgments
We thank John Cardy and Erwin Frey for valuable discussions on the corrections to scaling in the renormalisation group.
Financial support by the Nanosystems Initiative Munich (NIM) is gratefully acknowledged.

% \bibliographystyle{eplbib}
% \bibliography{percolation}

\end{document}